# Current Oscillations in Quasi-2D Charge-Density-Wave 1$T$-TaS$_2$ Devices: Revisiting the "Narrow Band Noise" Concept

Adane K. Geremew[1], Sergey Rumyantsev[1,2], Roger Lake[3] and Alexander A. Balandin[1,*]

[1]Nano-Device Laboratory (NDL) and Phonon Optimized Engineered Materials (POEM) Center, Department of Electrical and Computer Engineering, University of California, Riverside, California 92521 USA

[2]Center for Terahertz Research and Applications (CENTERA), Institute of High-Pressure Physics, Polish Academy of Sciences, Warsaw 01-142, Poland

[3]Laboratory for Terascale and Terahertz Electronics (LATTE), Department of Electrical and Computer Engineering, University of California, Riverside, California 92521, USA

## Abstract

We report on current oscillations in quasi two-dimensional (2D) 1$T$-TaS$_2$ charge-density-wave devices. The MHz-frequency range of the oscillations and the *linear* dependence of the frequency of the oscillations on the current resemble closely the "narrow band noise," which was often observed in the *classical* bulk quasi-one-dimensional (1D) trichalcogenide charge-density-wave materials. In bulk quasi-1D materials, the "narrow band noise" was interpreted as a *direct* evidence of the charge-density-wave sliding. Despite the similarities, we argue that the nature of the MHz oscillations in quasi-2D 1$T$-TaS$_2$ is different from the "narrow band noise." Analysis of the biasing conditions and current indicate that the observed oscillations are related to the current instabilities due to the voltage-induced transition from the nearly commensurate to incommensurate charge-density-wave phase.

**Keywords:** narrow band noise, 1T-TaS$_2$, charge density waves, hysteresis

---

[*] Corresponding author (A.A.B.): balandin@ece.ucr.edu ; web-site: http://balandingroup.ucr.edu/

It is well known since the mid-20th century that some materials, usually metals, with quasi one-dimensional (1D) crystal structure can form the charge-density-wave (CDW) state [1]. Examples of such materials, many of them transition metal trichalcogenides (TMTs), include $NbS_3$ and $TaS_3$. A large number of interesting CDW-related effects have been observed experimentally and studied theoretically [1-9]. They include strongly nonlinear electrical conductivity, phase transitions induced by an electric field, de-pinning and sliding of CDWs under applied electric bias. One of the most interesting phenomena observed for quasi-1D bulk CDW materials was the "narrow band noise" [2-9]. It was found that at the bias voltage, which corresponds to a certain electric field, exceeding the threshold field, $E_T$, the noise spectrum at MHz frequency range reveals one or several well-defined peaks. The effect was associated with the de-pinning and sliding of the CDWs. The peaks in the noise spectra were interpreted as an indication of the periodic current oscillations due to collective current of the sliding CDW. The term "noise" in this context is not quite correct. The effect should have been called the current oscillations. However, it has been used extensively in literature, and we mention it here for a historic prospect.

The studies of the current oscillations, *i.e.* "narrow band noise", in bulk 1D CDW materials observed a linear relationship between its frequency, $f$, and the collective, *i.e.* excess, current, $I_{CDW}=I-I_N$, attributed to the CDW sliding, where $I$ is the total measured current and $I_N$ is the conventional current due to individual electrons, extracted from Ohm's law. The linear relationship was explained by the assumption that $f$ is proportional to the CDW drift velocity, $v_D$, so that $f=v_D/\Lambda$, where $\Lambda$ is some characteristic distance. Since $I_{CDW}=nef\Lambda A$, where $n$ is the charge carrier density, $e$ is the charge of an electron, and $A$ is the cross-sectional area of the sample, one can write for the frequency $f=(1/ne\Lambda A)\times I_{CDW}$ [1]. Other models, with more material specificity, resulted in a similar relationship between the frequency and current. The observation of the "narrow band noise" was interpreted as the direct evidence of the CDW current, which arises from the sliding of CDWs. The "narrow band noise" was also related to a special new type of instability and negative differential dielectric constant.

Recent years witness a rebirth of the CDW field [10-20]. It has been associated, from one side, with the interest to layered quasi-two-dimensional (2D) van der Waals materials and, from another side, with realization that some of these materials reveal CDW effects at room temperature (RT) and above [16]. Examples of these materials, many of them transition metal dichalcogenides (TMDs), include $1T$ polymorph of $TaS_2$ and $1T$ polymorph of $TaSe_2$. The most interesting 2D layered TMD with CDW properties is the $1T$ polytype of $TaS_2$. It undergoes a transition from a normal metallic phase to an incommensurate CDW (IC-CDW) phase at 545 K, then to a nearly-commensurate CDW (NC-CDW) phase at 350 K, and, finally, to a commensurate CDW (C-CDW) phase at ~200 K [10-16]. The C-CDW phase consists of a $\sqrt{13} \times \sqrt{13}$-reconstruction within the basal plane that forms a star-like pattern, in which each star contains 13 Ta atoms. The Fermi surface, composed of 1d-electron per star, is unstable. The lattice reconstruction is accompanied by a Mott-Hubbard transition that gaps the Fermi surface and increases the resistance.

As the temperature increases, the C-CDW phase breaks up into the NC-CDW phase, which consists of the C-CDW domains. The C-CDW to NC-CDW transition is revealed as an abrupt change in the resistance with a large hysteresis window in the resistance profile at ~200 K. As the temperature is increased to ~350 K, the NC-CDW phase melts into the I-CDW phase. This transition is accompanied by a smaller hysteresis window in the resistivity. It has been conformed in numerous studies that $1T$-$TaS_2$ -- quasi-2D CDW quantum material -- reveals numerous intriguing properties, including multiple phase transitions, accompanied by resistance change and hysteresis, dependence of the phase transition points on the number of atomic planes, and "hidden" phases at low temperature [10-16]. Intercalation, doping, pressure, optical excitation, and electric source-drain bias can modify the transitions to different CDW phases in $1T$-$TaS_2$. The demonstrated RT operation of CDW oscillators [16], designs of the transistor-less circuits on the basis of CDW devices [17-18], and intriguing radiation hardness against X-ray and protons [19-20] transitioned the 2D CDW research field into the domain of applied physics and practical applications.

Despite the strong interest to $1T$-$TaS_2$ and other quasi-2D the "narrow band noise" effect has never been observed in such materials. The *current fluctuations*, which were interpreted as a direct evidence of CDW de-pinning and sliding in "classical" bulk quasi-1D CDW materials [1-9], for some reason are missing in the "new generation" quasi-2D CDW materials. In our previous studies of $1T$-TaS$_2$ we investigated the low-frequency noise, termed the "broad band noise" in the CDW community [1], conducting measurements from 1 Hz to ~100 kHz [21-23]. We have observed peaks in the noise spectral density as a function of bias at the frequencies below 100 kHz, which were related to the CDW phase transitions, de-pinning and possible "hidden phases" [21-23]. However, no current fluctuations, *i.e.* "narrow band noise" in MHz or other spectral range, have ever been reported for $1T$-TaS$_2$ or other quasi-2D CDW materials. In this Letter, we report an observation of the current fluctuations in $1T$-TaS$_2$ at MHz frequency range, which reveal striking similarity to the "narrow band noise" in "conventional" bulk quasi-1D CDW materials such as NbS$_3$ and other CDW TMTs. Unexpectedly, we found that the noise peaks in quasi-2D CDW test structures can be explained by the hysteresis at the IC-CDW – NC-CDW phase transitions rather than by sliding of CDWs. We offer a possible explanation for the absence of the *true* "narrow band noise" in quasi-2D CDW materials.

Thin films of crystalline $1T$-TaS$_2$ were mechanically exfoliated and transferred to SiO$_2$/Si substrates. The thickness of SiO$_2$ layer was always 300 nm. The metal Ti/Au contacts were evaporated through a shadow mask to avoid chemical contamination and oxidation during the fabrication process. The shadow masks were fabricated using double-side polished Si wafers with 3 μm thermally grown SiO$_2$ (Ultrasil Corp.; 500-μm thickness; P-type; <100>). For the devices used in the study, the $1T$-TaS$_2$ thickness was in the range from 10 nm to 5,0 nm, the channel length was about 2 μm, and the width ranged from 10 μm to 15 μm. Details of fabrication of $1T$-TaS$_2$ two-terminal devices, *i.e.* test structures, have been reported by us elsewhere [21-22]. The current-voltage ($I$-$V$) characteristics and resistivities were measured with a semiconductor analyzer (Agilent B1500). The noise measurements were conducted under DC bias voltages, $V_{SD}$, in the frequency range from 50 kHz to 2 MHz. The voltage fluctuations from the 1 kΩ load resistor were amplified by low noise amplifier (Stanford Research SR 560) and analyzed by spectrum analyzer (Brüel and Kjær). The value of the load resistance was selected to be close to that of the device under test as normally done in the noise measurements. Since the nominal bandwidth of the amplifier (SR 560) is 1 MHz, the gain at higher frequencies was measured separately as a function of frequency, and the actual gain at given frequency was used to plot the spectra. More information on our noise testing protocols can be found in prior publications [21-26].

Figure 1 shows the dependence of noise spectral density, $S_I$, on the current through the device channel at frequency $f$=760 kHz. To ensure the reproducibility the data are shown for two test structures. The general trend is $S_I \propto I^2$, as one typically observes in the low-frequency noise measurements for various materials [27-28]. However, one can notice some deviation from the $I^2$ scaling, *i.e.* steeper increase in $S_I$ with the current. At low current levels, $I < 8 \times 10^{-4}$ A, the noise spectral density, $S_I$, had the form of the $1/f$ noise, *i.e.*, $S_I(f) \propto 1/f$. As the current increased, the spectral density became flatter, possibly indicating the increasing contribution of the generation-recombination-type noise, which has Lorentzian spectrum. This spectrum flattening is likely responsible for $S_I$ deviation from the $I^2$ scaling. The insets to Figure 1 show the device schematic and the normalized dependence of the gain on frequency.

[Figure 1]

The measurements of the noise spectral density, $S_I$, as the function of frequency, $f$, at higher currents, $I > 2$ mA, demonstrated the peak, closely resembling the "narrow band noise" observed in bulk quasi-1D CDW materials [1-9]. Figure 2 shows the "noise" spectra close to the peak for a representative device at different currents. The spectra were obtained by normalizing the measured spectra by the amplitude – frequency characteristic shown in the inset to Figure 1. As one can see, the peak shifts with the current increase like the current dependence of the "narrow band noise," reported in numerous publications for bulk samples of quasi-1D CDW materials [1-9]. Every peak in Figure 2 corresponds to the periodic current oscillations with the fundamental frequency $f_{max}$.

[Figure 2]

The dependence of the current oscillation frequency, $f_{max}$, on the current is presented in Figure 3. The extracted current dependence of the oscillation frequency is close to linear. As we discussed above, the frequency of the "narrow band noise" depends linear on the current according to different models, *i.e.*, $f=(1/ne\Lambda A)\times I_{CDW}$ [1-9]. In this sense, the observed noise peaks in $1T$-TaS$_2$ devices in MHz frequency range closely resemble the "narrow band noise" in the bulk quasi-1D CDW materials such as NbS$_3$. However, there is one substantial difference with the classical "narrow band noise" experiments. The current density in our experiments, at which the peaks are observed, is on the order of $10^6$ A/cm$^2$, which is many orders of magnitude larger than in "narrow band noise" experiments with bulk quasi-1D CDW materials [1]. For this reason, the mechanism of the current oscillations in quasi-2D $1T$-TaS$_2$ might be different.

We hypothesized that the current oscillations of a particular frequency, $f_{max}$, in the MHz range can be not the signatures of the CDW sliding but of the IC-CDW – NC-CDW phase transition, similar to the one used by the present authors for demonstration of the CDW voltage-controlled oscillator [16]. Figure 4 shows the *I-V* characteristic of the same $1T$-TaS$_2$ device used to demonstrate the "noise" peaks (Figure 2). The dashed line represents the load for 1 kΩ resistor, and for one of the total applied voltages used in the experiments. The hysteresis loop with the on-set at 0.8 V is the phase transition from NC-CDW to IC-CDW induced by the applied electric bias to the device channel. This type of the phase transition is well known, and it has been reported in various independent studies, including by the means of the noise spectroscopy [12, 16, 21]. At this bias, the circuit has two working points at different voltages and currents. The state with the smaller voltage on the sample corresponds to the higher current. Although the details of phase transition dynamic are not well known [12], we speculate that the observed current oscillations relate to the instability at the voltages close the phase transition hysteresis loop. The oscillation mechanism can be understood as the switching of the $1T$-TaS$_2$ device between two conducting states with the negative feedback from the load resistor [16]. This mechanism is different from the "narrow band noise" observed in quasi-1D bulk CDW materials, where it was attributed to the CDW sliding [1].

The scaling of the frequency of the oscillations with the current for mechanism based on NC-CDW – IC-CDW phase transition cannot be established accurately at this point. It requires the detailed knowledge of the current dynamics close and inside the hysteresis loop. The frequency can also be affected by the parasitic elements of the device test structure and measurement setup. For a rough estimate one can use the circuit model of $1T$-TaS$_2$ device, proposed in Ref. [16]. It describes the frequency, $f=1/T$, determined by the charging and discharging time of the lumped capacitance between the device output node and the ground. If we assume that the change in the bias voltage, $V_{DC}$, applied across the device channel and the load resistor, is small compared to $V_{DC}$ near the hysteresis, one can write, in the linear response, $\delta f = (-1/T^2)(dT/dV_{DC})\delta V_{DC}$. The latter suggests a linear increase in frequency with the current through the device over the hysteresis window range of voltages. However, further investigation is needed to verify this conclusion.

The intriguing question, which remains, is why nobody observed the true "narrow band noise" in quasi-2D CDW TMD materials such as $1T$-TaS$_2$. One possibility can be that in devices operating in the NC-CDW phase, the collective CDW current conduction is always overshadowed by the conventional individual electron conduction. It has been noted in prior studies that the onset of non-linear conduction in 1T-TaS$_2$ is much less pronounced than in bulk quasi-1D CDW materials [21-22]. Another possibility can be that the characteristic distance, $\Lambda$, in the equation for the current oscillation, $f=(1/ne\Lambda A)\times I_{CDW}$, which can be associated with the CDW coherence length is shorter, and much broader distributed, in 2D layered material than in 1D layered materials. The third, most drastic possibility, which would require systematic re-investigation of CDW effects in bulk quasi-1D TMT materials, is that the presence of hysteresis may have been overlooked at least in some earlier studies. The CDW de-pinning is always accompanied by the nonlinear conduction and, in some cases, by hysteresis effects [1-6]. This hysteresis close to the voltage of the CDW de-pinning can lead to the oscillations, similar to the situation observed in the present study. The "narrow band noise" was typically observed using noise methods, and revealed itself as oscillations just above the background noise. This means that the "narrow band" oscillations are always of the small amplitude. Similar oscillations can be observed just due to the hysteresis at the offset to de-pinning and not by the sliding itself. Since the amplitude of the oscillations is small, the small hysteresis, which causes these oscillations could have been overlooked in the *I-V* characteristics measurements.

In summary, being puzzled by the absence of the reports of the "narrow band noise" in quasi-2D CDW materials, we undertook investigation of the noise response of 1T-TaS$_2$ thin-film test structure to the MHz frequency range. At certain values of the current through the two-terminal devices, we found the current oscillations, with the frequency proportional to the value of the current, which resemble closely the classical "narrow band noise" bulk quasi-1D trichalcogenide CDW materials. However, despite the similarities, we argued that the nature of the MHz oscillations in 1T-TaS$_2$ is different, and it is related to the current instabilities at the NC-CDW to IC-CDW transition rather than to the CDW sliding. The obtained results are important for better understanding of electron transport and phase transitions in quasi-2D CDW materials and for their proposed device applications.

**Acknowledgements:** A.A.B. acknowledges support by the National Science Foundation (NSF) through the Designing Materials to Revolutionize and Engineer our Future (DMREF) Program entitled Collaborative Research: Data Driven Discovery of Synthesis Pathways and Distinguishing Electronic Phenomena of 1D van der Waals Bonded Solids, and by the UC-National Laboratory Collaborative Research and Training Program - University of California Research Initiatives LFR-17-477237. The work of S.R. was partially supported by CENTERA project carried out within the IRAP program of the Foundation for Polish Science (grant MAB/2018/9) co-financed from the EU European Regional Development Fund. Nanofabrication has been performed in the UC Riverside Nanofabrication Facility.

**Contributions:** A.A.B. and S.R. conceived the idea. A.B. coordinated the project, led the data analysis, and manuscript preparation; A.K.G. fabricated the devices, conducted electrical measurements, and assisted with the noise measurements. S.R. conducted the noise measurements and data analysis; R.L. contributed to data analysis. All authors contributed to the manuscript preparation.

**FIGURE CAPTIONS**

**Figure 1**: Noise power spectral density, $S_I$, as a function of the current through 1T-TaS$_2$ device channel measured at frequency $f$=760 kHz. The red and blue data points correspond to two tested devices. The upper inset shows a schematic of the two-terminal CDW devices; the lower inset shows the gain, normalized to the gain at $f$=30 kHz, as a function of frequency.

**Figure 2**: Noise power spectral density as a function of frequency for several value of the current through the device channel. The spectra were obtained by normalizing the measured spectra by the amplitude-frequency characteristic shown in the inset in Figure 1. The peak shifts to the higher frequency $f_{max}$ with the increasing current, closely resembling the "narrow band noise" observed for bulk samples of quasi-1D CDW materials. The inset shows the schematic of the measurement.

**Figure 3**: Frequency, $f_{max}$, of the noise peaks as a function of the current through 1T-TaS$_2$ device channel. The "narrow band noise" in bulk quasi-1D CDW materials reveled a linear dependence of the oscillation frequency on the current. The inset shows a microscopy image of a representative 1T-TaS$_2$ device structure with several metal contacts.

**Figure 4:** Current-voltage characteristics of tested 1T-TaS$_2$ device (the same as in Figure 2). The hysteresis loop at the bias voltage V = 0.9 V corresponds to the transition from the NC-CDW phase to the IC-CDW phase induced the applied electric field. The arrows indicate the forward and reverse cycle currents.

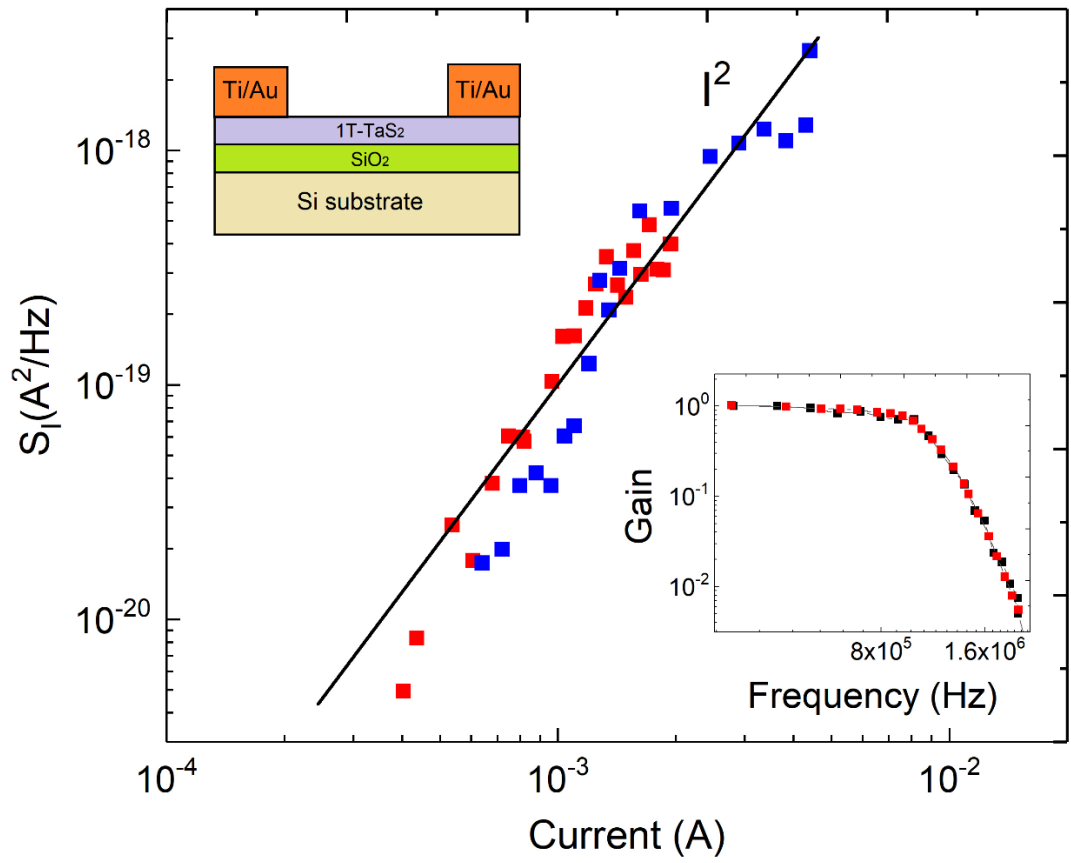

Figure 1

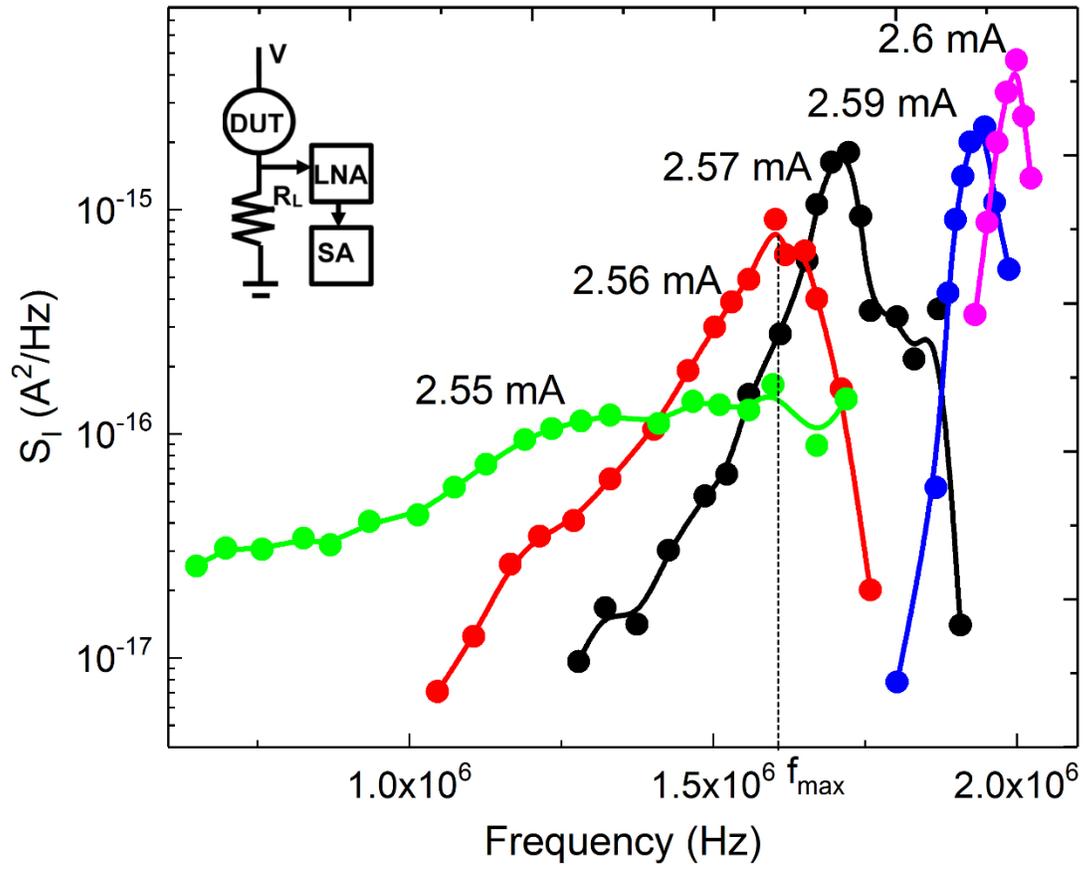

Figure 2

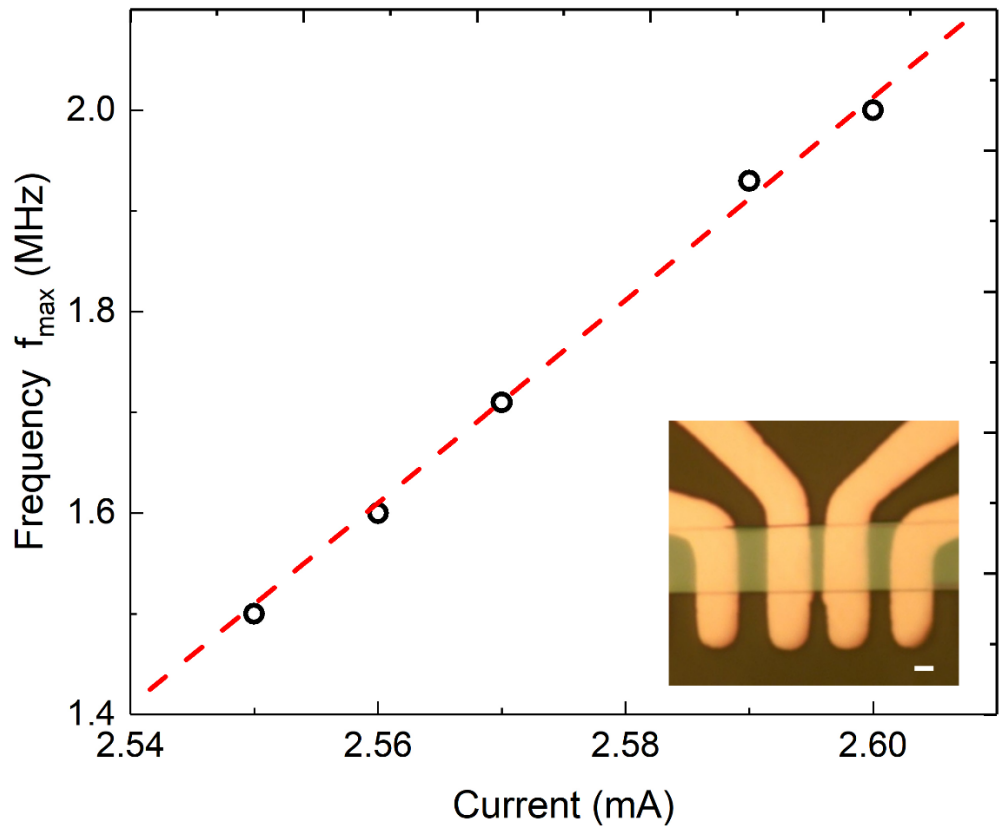

Figure 3

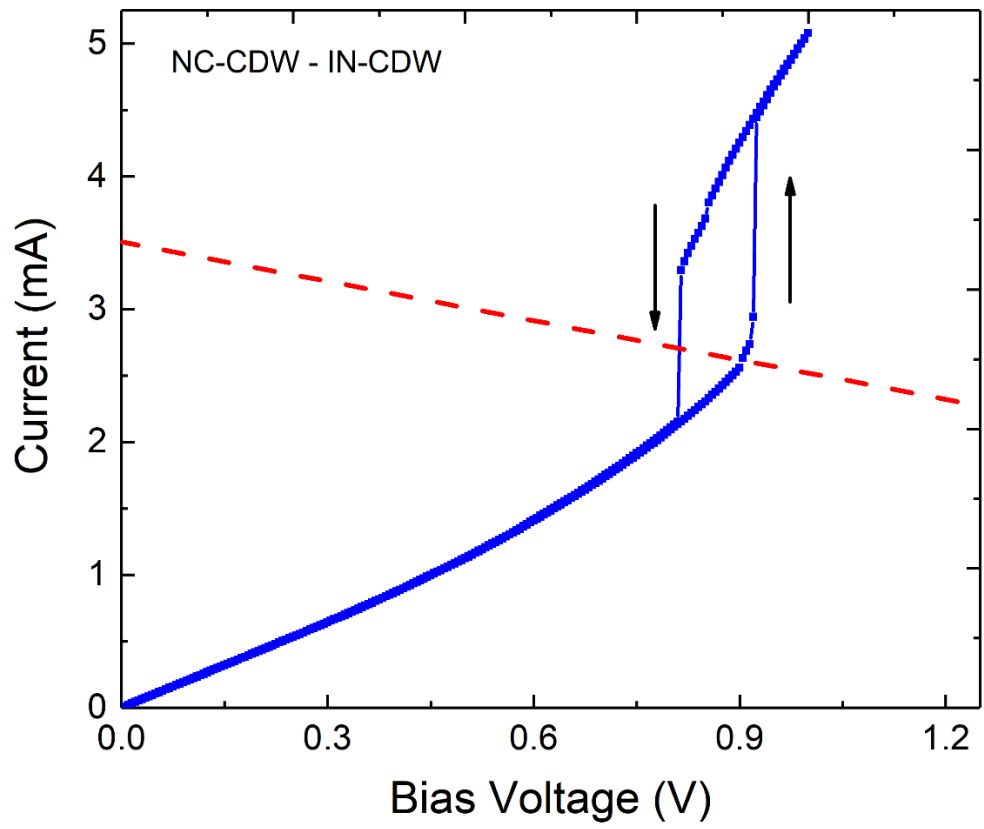

Figure 4